\documentclass[draft]{revtex4}
\usepackage{graphicx}

\begin{document}
\markboth{R. S. Cara\c ca and Manuel Malheiro}{Charge Influence On Mini Black Hole's Cross Section}

\title{CHARGE INFLUENCE ON MINI BLACK HOLE'S CROSS SECTION}

\author{ R. S. CARA\c CA* AND M. MALHEIRO**}
\address{* Universidade de S\~ao Paulo, Instituto de F\'isica, \\
    05314-970, S\~ao Paulo, Brazil,\\
    **Instituto Tecnol\'ogico de Aeron\'autica, \\
             CTA, 12228-900, S\~ao Jos\'e dos Campos, Brazil
             \\
         rcaraca@usp.br ,malheiro@ita.br }

\begin{abstract}
In this work we study the electric charge effect on the cross section production
 of charged mini black holes (MBH) in accelerators. We analyze the charged MBH
  solution using the {\it fat brane} approximation in the context of the ADD model.
  The maximum charge-mass ratio condition for the existence of a horizon radius is discussed.
  We show that the electric charge causes a decrease in this radius and, consequently,
  in the cross section. This reduction is negligible for protons and light ions but can be
   important for heavy ions.
\end{abstract}

\maketitle

\section{Introduction}\indent\indent

In 1998, Nima Arkani-Hamed, Savas Dimopoulos and Gia Dvali, introduced a novel approach to the hierarchy problem
 between gravitation, which characteristic scale is the Planck scale, $M_P\sim 10^{16}$ TeV,
  and the Standard Model which has an energy scale $M_{SM}\sim 1$ TeV.\par
This model is based on the existence of branes and large extra
dimensions (LED)\cite{add98}.
 According to this, our four dimensional world is a 3-brane, embedded in a higher dimensional manifold,
  called {\it bulk}, but the gauge fields are trapped on the 3-brane and
   only  gravitation has access to all dimensions of spacetime so that its intensity is diluted across the extra dimensions,  explaining the weakness of  gravity when compared with the other fundamentals interactions of nature. This model is known as the ADD (Arkani, Dimopoulos, Dvali).\par
The fundamental scale of gravitation, in ADD model, is considered to be on the order of $M_D\sim$ 1 TeV, while the value measured (effective) in our four dimensional world is the Planck scale: $M_P\sim 10^{16}$ TeV.\par
In the brane models there is a relation between the Planck scale in four dimensions ($M_P$) and the Planck scale in $D$-dimensions ($M_D$)\cite{add98,kanti}:\\
$$M_P^2\sim M_D^{D-2}R^{D-4},$$\\
where $R$ is the characteristic radius of the LED and  ranges from $R\sim 0.1mm$ up to $R\sim 1fm$.
 Moreover brane models are based on (low energy limit) string theories so, usually, the number of spacetime dimensions, $D$,  goes up to 11.\par
One characteristic of the ADD model is to provide a possible explanation for hyerarchy problem between the gravitational constant, $G_4=M_P^{-2}$ in our four dimensional spacetime and  $G_D=M_D^{2-D}$ in higher dimensional brane worlds.
$$\frac{G_D}{G_4}\sim \left(\frac{M_P}{M_D}\right)^2\frac{1}{M_D^{D-4}},$$\\
where $M_P/M_D\sim 10^{16}$ with $M_D\sim 1$ TeV.\par Another
consequence of the ADD model is the possibility to create mini
black holes (MBH) in particle colliders, such as the LHC. This
fascinating hypothesis has arose the interest about the matter,
what can be seen, for example, on
references\cite{kanti,stocker1,stocker07,stocker08,hossenfelder,hossenfelder3,giddings}.

\section{The fat brane approximation}\indent\indent
In theory, to produce a MBH, we need energies above the fundamental scale, which  in our four dimensional world
 is $M_P\sim 10^{16}$ TeV, but it is totally inaccessible to the experiments. However, in the ADD scenario, MBH could be produced with energies of approximately some TeV.\par
The Schwarzschild-Tangherlini solution\cite{tangh} in ADD model
produces the following expression
 for the horizon radius\cite{kanti}
\\
$$r_S=\frac{1}{\sqrt{\pi}}\frac{1}{M_D}\left(\frac{8\Gamma\left(\frac{D-1}{2}\right)}{D-2}\right)^{1/(D-3)}
\left(\frac{M}{M_D}\right)^{1/(D-3)},$$
\\
where $\Gamma\left(\frac{D-1}{2}\right)$ is the Gamma function and is related to area of an hypersphere
 of $D$ dimensions: $A_D=\frac{2\pi^{(D-1)/2}}{\Gamma((D-1)/2)}$.\par
In this paper we will to estimate the electric charge effects on the cross section of MBH.
In order to do that, we need to confine the electric field in the brane and find the expression
 for $D$-dimensional charged black holes, however, a exact solution for a compactified  electric field is complicated and not known up to now. A complete treatment needs  to consider effects of backreaction, because the electromagnetic fields (and so the gauge fields) gravitate due to a ``effective mass'' that is associated to it.\par
  We do not treat in this work effects of backreaction. Our aims are to get a estimated charge influence on the MBH production using the the {\it fat brane} approximation. Recently, some aspects of charge and mass on evaporation of MBH considering backreaction effects were discussed on reference\cite{sampaio}. Another discussion about charged objects in extra dimensions is treated in\cite{vilson,lemos}. However, in these last two references the authors consider a spacetime with $D$ extended dimensions (non compactified).\par
Roughly, the {\it fat brane} approximation consists to give a
nonzero thickness for the brane and up to distances $\lesssim
\mbox{TeV}^{-1}$ we assume that all gauge fields can penetrate in
the {\it bulk}\cite{hossenfelder2}. In this approximation we found
a solution that locally is the same as that obtained by Myers \&
Perry\cite{myers}
  for $D$-dimensional charged black holes with a mass $M$ and charge $Q_D$:\\
\begin{eqnarray}\nonumber
    ds^2&=&-\left(1-\frac{16\pi G_DM}{A_D(D-2)r^{D-3}}+\frac{2 G_D  Q_D^2}{(D-2)(D-3)}\frac{1}{r^{2(D-3)}}\right)dt^2+\\
    &&+\frac{1}{\left(1-\frac{16\pi G_DM}{A_D(D-2)r^{D-3}}+\frac{2 G_D  Q_D^2}{(D-2)(D-3)}\frac{1}{r^{2(D-3)}}\right)}dr^2+r^2dA_D^2.
\end{eqnarray}\\ \par
In the metric above we have $g_{tt}(r)=g_{rr}^{-1}(r)$ and when we take $r\to r_H$, where:
\begin{eqnarray}\label{rnradius}
r_H&=&\left[\frac{4 G_D\Gamma[(D-1)/2]M}{\pi^{(D-3)/2}(D-2)}\left(1+\sqrt{1-\frac{Q_D^2\pi^{D-3}(D-2)}{8G_D[\Gamma((D-1)/2)]^2(D-3)M^2}}\right)\right]^{1/(D-3)},
\end{eqnarray}
the coefficients $g_{tt}(r)=0$ and $g_{rr}(r_H)\to \infty$, that
indicate the presence of an event horizon,
 in analogy with Reissner-N\"{o}rdstron case in four dimensions.\par
If the expression inside the square-root in (\ref{rnradius})
vanishes, we have a extremal case,
 analogue to extremal black holes in four dimensions. This condition gives the maximal charge that allows
 the event horizon formation. If, otherwise, this expression assumes a negative value, we do not to expect
  neither the event horizon formation and nor a black hole, in according with cosmic censorship conjecture\cite{wald}.\par
In our case, for the existence of MBH horizon the term in the
square-root needs to be non-negative and we have the following
charge-mass condition
\begin{eqnarray}\label{qm}
Q_D^2&\leq&\frac{8\left[\Gamma\left(\frac{D-1}{2}\right)\right]^2(D-3)}{\pi^{D-3}(D-2)}G_DM^2,
\end{eqnarray}\\
where $Q_D$ is the electric charge in $D$-dimensions and $M$ is the MBH mass.\par
The equation (\ref{qm}), up to a geometric factor, has the same form of the charge-mass
 condition in four dimensions, $Q_4^2\leq (\sqrt{G_4}M)^2$, but now in $D$-dimensions.
  It is important to stress that the charge ($Q_D$) depends on the spacetime dimensions
    since $Q_D$ is proportional to $\sqrt{G_D}$ and that varies with the dimensions $D$.

\section{Charge Influence on Cross Section of MBH}\indent\indent

 Accelerators such as the LHC will collide charged particles so it is important
  to know how the electric charge will affect the cross section and will change the MBH production.\par
A complete description of MBH physics needs a quantum gravity theory, not yet developed.
 However, when the MBH mass is larger than the fundamental scale $M_D$,
   we can use a semiclassical approach to approximate the cross section for a charged MBH as\cite{giddings}\\
\begin{eqnarray}\label{eq3}
    \sigma_{ij}&\approx& \pi r_H^2,
\end{eqnarray}\\
where $r_H$ is the horizon radius given by (\ref{rnradius}).\par
In Table 1, we present the values for the maximum charge-mass
ratio
 $Q_D^2/(\sqrt{G_D}M)^2$ for different spacetime dimensions obtained from (\ref{qm}).\par
\begin{table}[!h]
    \centering
        \begin{tabular}{c|c|c|c|c|c|c|c}\hline
         &D=4   &D=6    &D=7    &D=8    &D=9    &D=10   &D=11\\\hline
        $\left(\frac{Q_D}{\sqrt{G_D}M}\right)_{max}^2$  &1.00 &0.34 &0.26   &0.24   &0.26   &0.31   &0.43\\\hline
        \end{tabular}
%        \caption{Maximum charge-mass ratio $Q_D^2/(\sqrt{G_D}M)^2$ for different spacetime dimensions.}
        \label{rnb}
\end{table}
{\small Table 1: Maximum charge-mass ratio $Q_D^2/(\sqrt{G_D}M)^2$
for different spacetime dimensions.}

\vspace{0.5cm}

We see from Table 1 that  the maximum charge-mass ratio has a
minimum value for $D=8$ and increases for $D>8$.
 This behavior is due to the fact that $Q_D^2/(\sqrt{G_D}M)^2$ is proportional to the inverse
  of the hypersphere area with unit radius, $A_D^{-1}=\left(\frac{2\pi^{(D-1)/2}}{\Gamma((D-1)/2)}\right)^{-1}$.
   The hypersphere area has a maximum value for $D=8$, decreasing for $D>8$.\par
We present in Table 2 the percentual cross section reduction for
charged MBH in comparison
 with the uncharged case $\sigma_0$, for several values of the charge-mass ratio $Q_D^2/(\sqrt{G_D}M)^2$
  and different space-time dimensions. As we see, when the charge-mass ratio increases the cross section decreases.
   This reduction is more pronounced when the value of the charge-mass ratio approaches its maximum value
    (Table 1), as we can see comparing the results on Table 2 for $ 7 \leq D\leq 9$
     when $Q_D^2/(\sqrt{G_D}M)^2=0.2$.\par

\begin{table}[!h]
    \centering
        \begin{tabular}{c|c|c|c|c|c|c}\hline
         &D=6   &D=7    &D=8    &D=9    &D=10   &D=11\\\hline
         $\left(\frac{\sigma_0-\sigma_{0.01}}{\sigma_{0}}\right)\times 100$ &0.5\%  &0.4\%  &0.4\%  &0.4\%  &0.3\%   &0.2\% \\
         $\left(\frac{\sigma_0-\sigma_{0.1}}{\sigma_{0}}\right)\times 100$      &5.2\%  &5.4\%  &5.0\%  &3.8\%  &2.5\%   &1.6\%\\
         $\left(\frac{\sigma_0-\sigma_{0.2}}{\sigma_{0}}\right)\times 100$   &12.2\%    &13.7\% &13.0\% &9.8\%  &6.2\%   &3.6\%\\\hline
        \end{tabular}
%    \caption{Percentual reduction of cross section, for $Q_D^2/(\sqrt{G_D}M)^2=0.01, 0.1$ and $0.2$, in comparison with the uncharged  Schwarzschild-Tangherlini's cross section $\sigma_0$.}
    \label{porcentagem}
\end{table}
{\small Table 2: Percentual reduction of cross section, for
$Q_D^2/(\sqrt{G_D}M)^2=0.01, 0.1$ and $0.2$, in comparison with
the uncharged  Schwarzschild-Tangherlini's cross section
$\sigma_0$.}

\vspace{0.5cm}

We also see a decrease of the cross section with the increase of the dimensions (for $D\geq 7$)
 and this effect is more important for large values of $Q_D^2/(\sqrt{G_D}M)^2$.\par

 From Table 2 we can conclude
  that for protons and light ions the reduction
   of the cross section in comparison with Schwarzschild-Tangherlini (uncharged)
    solution is small since the charge-mass ratio $\sim 10^{-3}$ is small\cite{hossenfelder2}
    (we assume that $\alpha_S$ does not change with the spacetime dimension since
     the electric fields only penetrate the {\it bulk} at $\sim \mbox{TeV}^{-1}$
      in the {\it fat brane} approximation). When the  charge-mass ratio approaches
       its  maximum value (expressed in Table 1),
        the charge effects become relevant causing  a reduction of the 0cross section above ten percent.

\section{Conclusion}\indent\indent
From the results shown in this paper we conclude that, in the {\it fat brane} scenario, the electric charge causes a decreasing in the horizon radius and, consequently, in the cross section. However, this reduction is negligible for protons and light ions
since the cross section is not very different from the case of the Schwarzschild-Tangherlini uncharged black hole.
For heavy ions, when the electric charge is close to the maximum charge-mass limit, the cross section reduction is large and charge effects are not so small. Finally, we emphasize that all conclusions in this work is based on semiclassical approach and  complete treatment needs, probably, to consider backreaction effects of gauge fields over the background.

\section*{Acknowledgments}
The authors acknowledge support from ITA (Instituto Tecnol\'ogico de
Aeron\'autica), CNPq, Capes/FCT Brazil-Portugal collaboration
project 183/07, and FAPESP thematic project 2007/03633-3.

\end{document}